\documentclass[tikz,11pt,english,a4paper]{article}%%
\newif\ifajp
\makeatletter
\@ifclassloaded{revtex4-1}
  {\ajptrue}   % if revtex4-1 is loaded
  {\ajpfalse}  % otherwise
\makeatother

\ifajp
    \usepackage{amsmath}
    \usepackage{amsfonts}
    \usepackage{amssymb}
    \usepackage{graphicx}
    \usepackage{caption}
    
    \DeclareCaptionLabelSeparator{normal-colon}{\normalfont : } 

    \usepackage{ragged2e}
    
    \makeatletter
    \let\old@raggedright\raggedright
    \renewcommand{\raggedright}{\justifying}
    \makeatother
    \renewcommand{\footnote}[2]{ (#2)}
\else
    \usepackage{jcappub}
    \usepackage{amsmath}
    \usepackage{amsfonts}
    \usepackage{amssymb}
    \usepackage{graphics}
    \usepackage{caption}

    \DeclareCaptionLabelSeparator{normal-colon}{\normalfont : } 

    \usepackage{float}
    \usepackage[utf8]{inputenc}
    \usepackage{lmodern}
    \usepackage{fontawesome5}
    
    \newcommand{\onlinecite}[1]{\cite{#1}}
    \let\latexfootnote\footnote
    \renewcommand{\footnote}[2]{\latexfootnote{#1}}
\fi

\usepackage[colorlinks=true
  ,urlcolor=blue
  ,anchorcolor=blue
  ,citecolor=blue
  ,filecolor=blue
  ,linkcolor=blue
  ,menucolor=blue
  ,pagecolor=blue
  ,linktocpage=true
  ,pdfproducer=medialab
  ,pdfa=true
]{hyperref}
\usepackage{listings}

\newcommand{\connect}{\textsc{connect}}
\newcommand{\class}{\textsc{class}}
\newcommand{\camb}{\textsc{camb}}

\begin{document}

\title{CosmoSlider: An educational tool for cosmology}
% In a long title you can use \\ to force a line break at a certain location.
\ifajp
    \author{Andreas Nygaard}
    \email{andreas@phys.au.dk}
    \email{andreas.hansen@uzh.ch}
    \affiliation{Department of Astrophysics, University of Zurich, CH-8057 Zurich, Switzerland}

    \author{Steen Hannestad}
    \email{steen@phys.au.dk}
    \author{Thomas Tram}
    \email{thomas.tram@phys.au.dk}
    \affiliation{Department of Physics and Astronomy, Aarhus University, DK-8000 Aarhus C, Denmark}
\else
    \author[a]{Andreas Nygaard}
    \author[b]{\hspace{-0.72em}, Steen Hannestad}
    \author[b]{\hspace{-0.62em}, and Thomas Tram}
    \affiliation[a]{Department of Astrophysics, University of Zurich, CH-8057 Zurich, Switzerland}
    \affiliation[b]{Department of Physics and Astronomy, Aarhus University, DK-8000 Aarhus C, Denmark}
    \emailAdd{andreas@phys.au.dk}
    \emailAdd{andreas.hansen@uzh.ch}
    \emailAdd{steen@phys.au.dk}
    \emailAdd{thomas.tram@phys.au.dk}
\fi

\date{\today}

\ifajp

    \begin{abstract}
    Understanding how cosmological parameters influence the cosmic microwave background (CMB) power spectra is a central component of modern cosmology education, but interactive exploration is often limited by computational cost or technical complexity. We present CosmoSlider, a lightweight visualization tool that enables real-time exploration of CMB power spectra as multiple cosmological parameters are varied simultaneously. The tool employs a neural-network emulator implemented using TensorFlow Lite, allowing rapid evaluation of spectra without relying on large grids of precomputed models or on-demand execution of Einstein–Boltzmann solvers. CosmoSlider is available both as an iOS application and as a web-based tool, making it accessible across platforms and suitable for use in classrooms, lectures, and self-guided study. By providing immediate visual feedback, CosmoSlider supports the development of intuition for the physical processes underlying CMB anisotropies and serves as a complementary resource to traditional theoretical instruction.
    \end{abstract}

    \maketitle

    \newpage
    \hrule
    \tableofcontents
    \medskip
    \hrule
\else
    \abstract{Understanding how cosmological parameters influence the cosmic microwave background (CMB) power spectra is a central component of modern cosmology education, but interactive exploration is often limited by computational cost or technical complexity. We present CosmoSlider, a lightweight visualization tool that enables real-time exploration of CMB power spectra as multiple cosmological parameters are varied simultaneously. The tool employs a neural-network emulator implemented using TensorFlow Lite, allowing rapid evaluation of spectra without relying on large grids of precomputed models or on-demand execution of Einstein–Boltzmann solvers. CosmoSlider is available both as an iOS application and as a web-based tool, making it accessible across platforms and suitable for use in classrooms, lectures, and self-guided study. By providing immediate visual feedback, CosmoSlider supports the development of intuition for the physical processes underlying CMB anisotropies and serves as a complementary resource to traditional theoretical instruction.}

    \maketitle
\fi

\section{Introduction} 
\begin{figure}[tb]
	\centering
	\includegraphics[width=\textwidth]{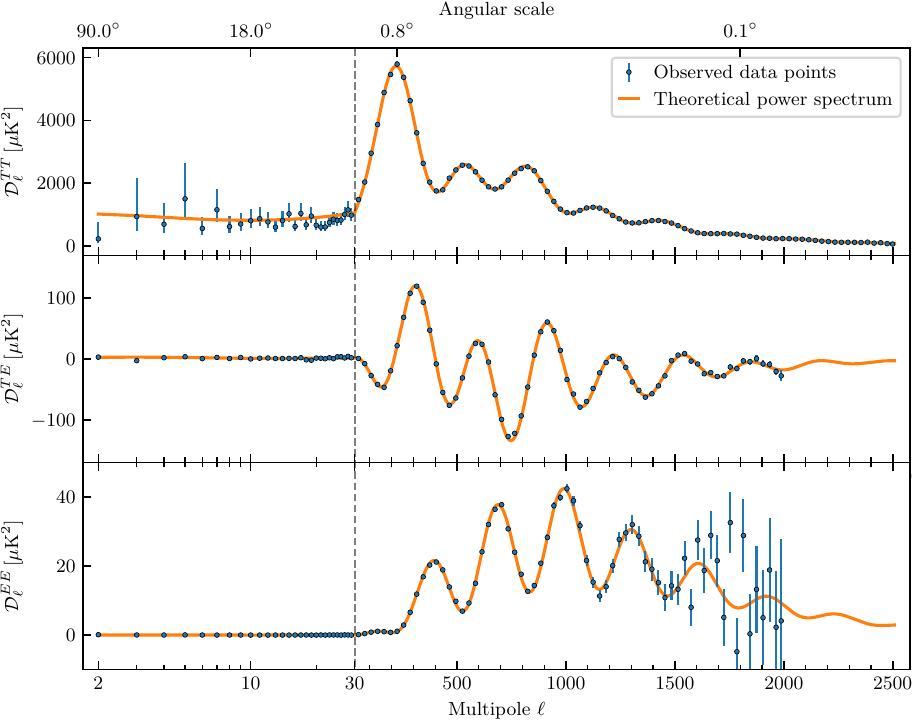}
	\caption{The CMB power spectra computed by \class{} using the Planck best-fit parameters~\cite{Planck:2018vyg} along with the observed data points from Planck.}
	\label{fig:cmb}
\end{figure}

When teaching new students about the early universe and how the cosmic microwave background (CMB) depends on specific cosmological parameters (such as the amount of dark matter, the expansion rate of the universe, etc.) lecturers and teaching assistants typically resort to some form of visualization of the CMB power spectrum as depicted in Fig.~\ref{fig:cmb}. This could be from actual computations using an Einstein--Boltzmann solver, such as \class{}~\cite{Blas:2011rf} or \camb{}~\cite{Lewis:1999bs}, or theoretical depictions motivated by the underlying physical phenomena (see Ref.~\onlinecite{waynehututorial} for examples hereof). When relying on actual computations to visualize the CMB power spectrum, a very popular method is creating a series of animations where a single parameter is changing (possibly with some underlying requirement of e.g. a flat universe). Publicly available examples of this include the CMB tutorials of Wayne Hu~\cite{waynehututorial} and a website by Max Tegmark~\cite{maxtegmarkuniverses}.

A more interactive approach than just having animations with a single varying parameter is allowing a user to change multiple parameters at once using e.g. a graphical slider element for each parameter. This of course is more demanding in terms of the number of precomputed spectra, since an $N$-dimensional grid is required for $N$ parameters, while only $N$ 1-dimensional grids are required when varying a single parameter at a time. With a resolution of $M=10$ spectra along each dimension and $N=6$ parameters, the interactive approach thus requires $M^N=10^6$ spectra while the animation approach only requires $M\times N=60$ spectra. Because of this much greater demand, publicly available tools using the interactive approach use either a small resolution, few parameters, or some kind of constraint to force correlation between sliders and bring down the number of required spectra. Publicly available examples of the interactive approach include the ``Build a Universe'' tool~\cite{buildauniverse}, originally created by the WMAP collaboration~\cite{wmap} and recreated for modern browsers by LAMBDA~\cite{nasalambda}, as well as a JavaScript visualizer by Jennifer Helsby and Eric Baxter~\cite{cosmowebapp}. The former uses 5 parameters with a resolution of 5 points per parameter direction and 1 parameter with a resolution of 4 points for a total of $12,\!500$ precomputed spectra, while the latter uses 6 parameters with some of them constrained to others (e.g. the three parameters $\Omega_m$, $\Omega_m h^2$, and $h$ are constrained to each other and $\Omega_{\Lambda}$ is also changed automatically to ensure flatness). The sliders of the Helsby--Baxter visualizer also reset to their default values when moving other sliders which makes the required number of precomputed spectra the same as for the animation approach but in turn makes the tool less versatile. Another tool utilizing interactive sliders is the CMB Simulator~\cite{cmbsimulator} created by Stuart Lowe and Chris North for a Planck exhibit which also shows a patch of the CMB sky map corresponding to the chosen parameters and features 3 sliders with a resolution of 41 for a total of $68,\!921$ precomputed spectra. The creators of the tool CMBverse~\cite{cmbverse} were able to use 6 sliders by dividing them into two separate groups, $\{\omega_{\Lambda}, \omega_b, \omega_{\rm cdm}\}$ and $\{\ln{\!(10^{10}A_s)}, n_s, \tau_{\rm reio}\}$. The first group has a resolution of 19 while the second has a resolution of 17 leading to a total number precomputed spectra of $11,\!772$. The website also features a few extra parameter plots with a single slider each with resolutions ranging from 30 to 101.

A way of avoiding numerous precomputed spectra while still maintaining a level of interactiveness is by updating the spectra only when prompted by the user. This way, the sliders or numerical input boxes can be set to the desired values before computing the corresponding spectra from scratch. LAMBDA hosts a web version~\cite{webcamb} of \camb{} with a user interface and a plethora of parameters. This is very versatile regarding the possible output, but in turn lacks the more interactive elements such as real time updates. A similar tool also relying on \camb{} to compute spectra on the fly is the CMB Explorer by Shivaji Chaulagain~\cite{cmbexplorer} which features the same functionalities as the web version of \camb{} but with a more modern user interface with sliders. Both of these tools require the user to press a ``Compute'' button in order for \camb{} to run, while a similar tool by Ross Jenkinson~\cite{rossjenkinson} runs \camb{} after a slider is released. This makes the user experience more interactive but requires a mandatory wait (a few seconds) after a slider is released since another slider cannot be moved while \camb{} is computing. This tool does not have as many parameters as the other \emph{compute-on-demand} tools and it only features the temperature power spectrum. Furthermore, it usually requires a warm up period of a few minutes due to inactivity.

\begin{figure}[tb]
	\centering
	\includegraphics[width=\textwidth]{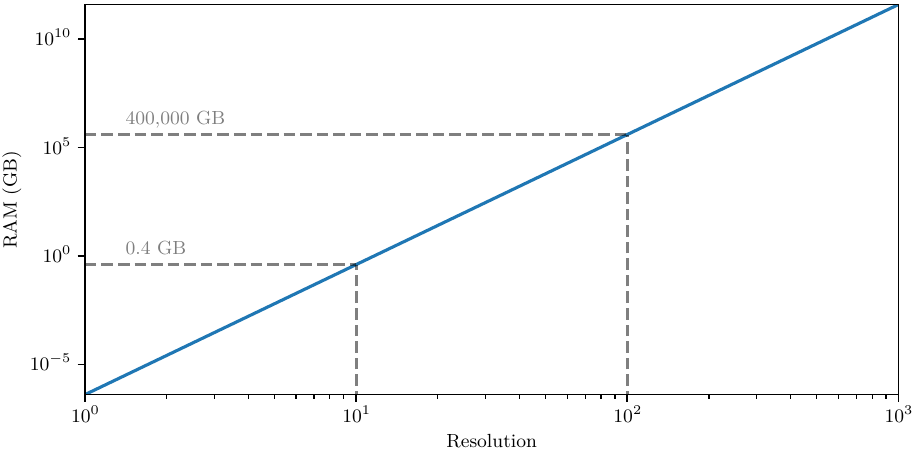}
	\caption{Amount of RAM needed to store $M^N$ precomputed spectra of 400 bytes as a function of the resolution, $M$, along each of $N=6$ dimensions. The dashed lines indicate the RAM needed specifically for resolutions of 10 and 100.}
	\label{fig:ram}
\end{figure}

A single CMB spectrum up to multipole $\ell = 2500$ can be well represented by computing 100 $C_\ell$ values across the range of $\ell$~\cite{Nygaard:2022wri}, and subsequent interpolation is quite fast if one deems it necessary when creating a visualization tool. This means that a single precomputed power spectrum (using 32-bit floating point numbers) takes up 400 bytes of memory, which puts a limit on the resolution and dimensionality when only relying on precomputed spectra. Fig.~\ref{fig:ram} shows how much RAM is needed by a tool storing the precomputed spectra necessary for a dimensionality of 6 (corresponding to $\Lambda$CDM) for different resolutions. From this, it is apparent that a tool solely relying on precomputed spectra cannot feasibly have a resolution much larger than $M=10$ for a dimensionality of $N=6$.   

One can of course bring down the number of precomputed spectra by interpolating between the spectra instead. This is in principle also done in the Helsby--Baxter visualizer, but only for one parameter at the time, since multiple parameters cannot be varied at once. Depending on the speed of the interpolator, a multidimensional interpolation could allow for a much higher resolution as well. The interpolation can be done either with standard interpolation algorithms or by training an emulator. The growing interest of machine learning within the field of cosmology has led to some very interesting applications. One in particular is the emulation of cosmological observables using artificial neural networks, and besides enabling a faster parameter inference, such an emulator can substitute the precomputed spectra needed for an interactive learning tool.

In this paper, we introduce the tool CosmoSlider which utilizes neural networks to emulate the power spectra. We describe the tool and its features in section~\ref{sec:cosmoslider}, elaborate on its potential use in educational settings in section~\ref{sec:education}, and sketch the future updates to the tool in section~\ref{sec:updates}. Finally, we draw our conclusions in section~\ref{sec:conclusion}.

\section{The CosmoSlider tool}\label{sec:cosmoslider}

There are various types of emulators to choose from, but in order to bring down the runtime of the emulation to a point where realtime updates are feasible, the emulator has to be lightweight, and to ensure versatility, it should be possible to easily train new emulators with different cosmological models. Thus, we use the \connect{}~\cite{Nygaard:2022wri} framework to create an accurate emulator that works well for the 6 parameters of the $\Lambda$CDM model. When using an emulator for realtime updates, the resolution of the graphical sliders can be arbitrarily high since no precomputed models are needed by the tool. The resolution would, however, pose a limit on the speed at which the sliders can be moved, since the emulator needs to update the spectrum for every step along the slider, but given an evaluation time of a single spectrum on the order of milliseconds, the resolution can easily be of order $M=10^2$ without affecting the user experience. It is also possible to ignore this problem and only update the spectrum when a slider is released (similar to Ref.~\onlinecite{rossjenkinson}), but this would not make for as smooth an interface.
\begin{figure}[tb]
	\centering
	\includegraphics[width=\textwidth]{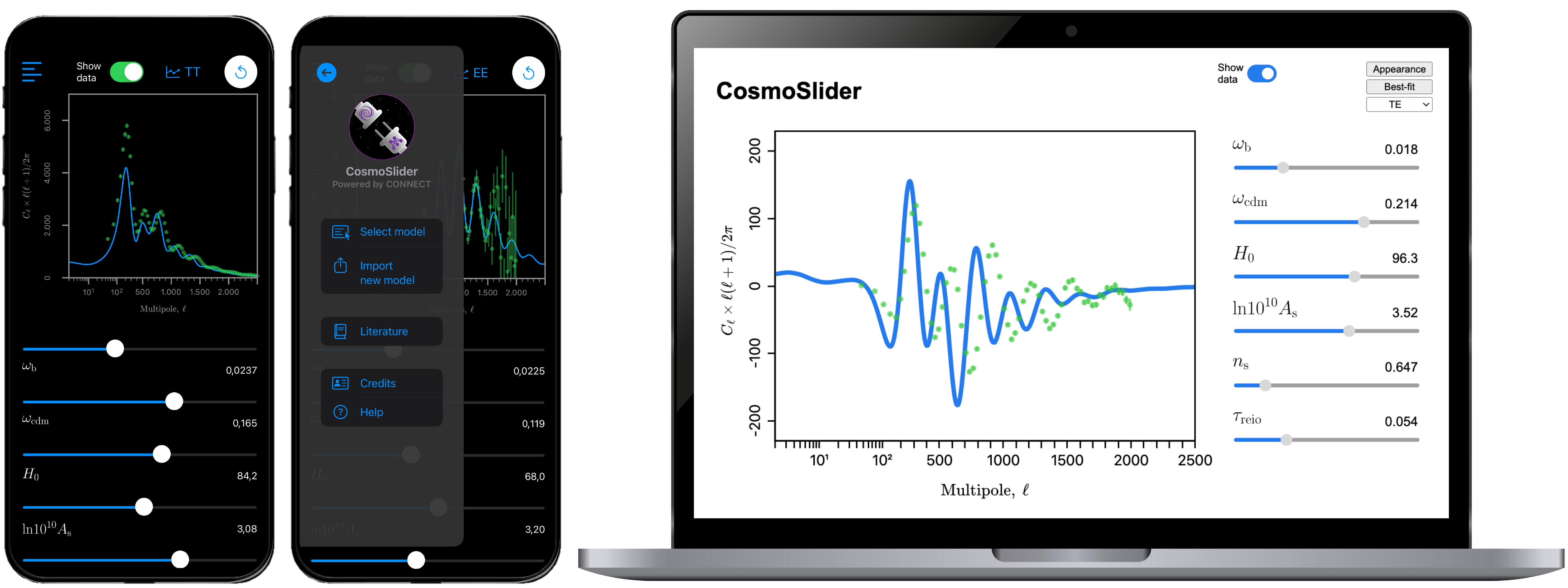}
	\caption{Visual representation of the two versions of CosmoSlider. On the left, the iOS version is shown on phone screens, and on the right, the web version is shown on a computer screen. Both versions have dark and light appearance modes.}
	\label{fig:screens}
\end{figure}

The CosmoSlider tool was initially created as an iOS application\footnote{Available for download here: \url{https://apps.apple.com/us/app/cosmoslider/id6504750261}}{available for download here: \url{https://apps.apple.com/us/app/cosmoslider/id6504750261}} using the SwiftUI language with a TensorFlowLite plugin in order to run the neural networks created by \connect{}, but in order to extend the reach to all users and make it more accessible to lecturers and teachers, we decided to create a web-based version\footnote{Accessible from here: \url{https://aarhuscosmology.github.io/CosmoSliderWeb}}{accessible from here: \url{https://aarhuscosmology.github.io/CosmoSliderWeb}} as well which uses the TensorFlowLite plugin for JavaScript. The two versions are depicted in Fig.~\ref{fig:screens}.

\subsection{Visualization of observables}

The observables are visualized using the native plotting utilities of SwiftUI and JavaScript with only few modifications. When displaying CMB power spectra, the plots switch from a logarithmic scaling for low multipoles to a linear scaling for high multipoles inspired by Ref.~\onlinecite{Efstathiou:2003dj} and this of course had to be implemented as it is not a standard feature of plotting utilities but offer a great way of examining both low and high multipoles in detail. Whenever a slider is being dragged, each step will update immediately ensuring smooth changes in the displayed spectra. With high resolutions of the different sliders ranging from $10^2$ to $10^3$, this requires very fast evaluation times provided by the neural network and the TensorFlowLite plugin for SwiftUI and JavaScript. In order to keep the evaluation time sufficiently short, only 100 $C_\ell$s are emulated, and the rest are obtained through cubic spline interpolation. The displayed spectrum type, i.e., TT, TE, EE, and $\phi\phi$, is easily switched using a drop-down menu in the top of the screen, where an option for resetting the sliders to their best-fit\footnote{According to the Planck Lite likelihood.}{according to the Planck Lite likelihood} positions can also be found. The user may also enable the visualization of data points from Planck~\cite{Planck:2018vyg} (SPT~\cite{Simard:2017xtw} for the $\phi\phi$ spectrum) which can guide in finding optimal values of the parameters.

\subsection{Features}
CosmoSlider is equipped with a few different features depending on the version. The web version currently doesn't have as many features as the iOS version, as it is a lighter implementation designed for a static webpage. This, however, makes it perfect for embedding on personal websites or course websites for educational use. The iOS application works on both iPhones and iPads running iOS 16.5 or higher and has a few extra features. This includes the possibility of changing the cosmological model and thus the varying parameters. It only ships preinstalled with the same $\Lambda$CDM model used by the web version, but it is possible to create a neural network using the \connect{} framework and import it as a new model in the iOS application. These models can then be shared with other users of the application and renamed at will. 
\begin{figure}[tb]
	\centering
	\includegraphics[width=\textwidth]{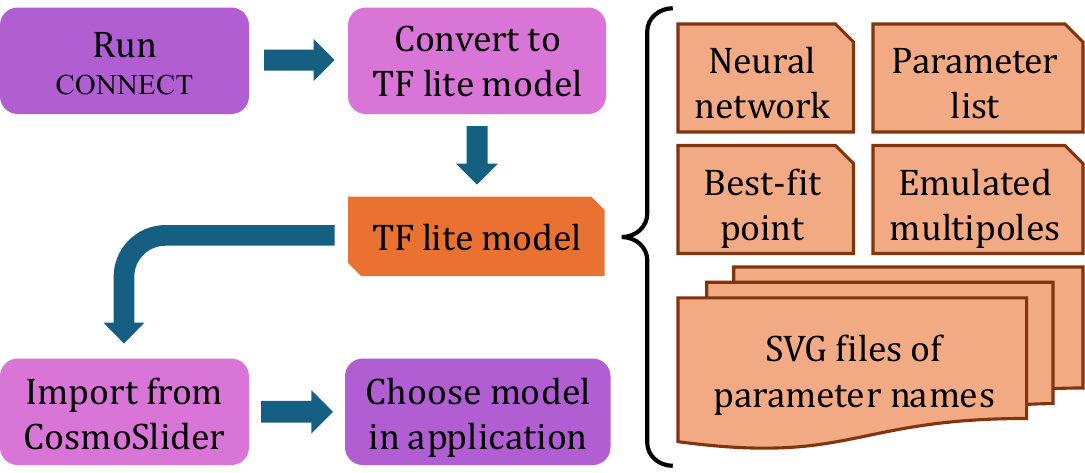}
	\caption{Flowchart of importing a new neural network model into the iOS version of CosmoSlider. The structure of the models read by the application is also depicted.}
	\label{fig:flowchart}
\end{figure}

In order to import a neural network from \connect{} into the application, the network needs to be converted into a TensorFlow Lite model. The details regarding the emulator, e.g., parameter names and ranges, the Planck best-fit point, and the emulated multipoles, also need to be extracted and put into files readable by the application. All of these including the TensorFlow Lite model, are gathered in a compressed \texttt{.tflite} file that can be imported into CosmoSlider. This procedure is depicted in Fig.~\ref{fig:flowchart} and is automatically done by \connect{} when prompted using the command
\begin{lstlisting}[language=bash]
  $ python connect.py convert /path/to/model
\end{lstlisting}
which creates a compressed \texttt{.tflite} file with the same name and location as the source model. This can then be transferred to a device with CosmoSlider installed and either opened with the application or imported from within it.

The iOS version also features a ``Help'' page along with a ``Litterature'' page for readers interested in either the CMB aspect or the emulation aspect. The CMB specific resources include guides to understanding the physical processes behind the CMB, while the emulation specific resources include various papers explaining the \connect{} framework.

\section{Educational usage}\label{sec:education}
CosmoSlider is developed as a visualization tool that can provide substantial benefits for both students and lecturers in educational settings. Lecturers can easily display the web version using a projector to demonstrate the effects of varying specific cosmological parameters, while students can download the iOS application or use the mobile web version to follow along and explore the parameter dependencies themselves, with no prerequisites beyond access to a smartphone. Because CMB power spectra can be generated in real time, CosmoSlider lowers the barrier for incorporating them into courses in more detail than they otherwise would have had the time to do with a more traditional approach involving Einstein--Boltzmann solvers.

\subsection{Building intuition}
Every student taking a course in astrophysics will eventually be introduced to the angular power spectrum of the cosmic microwave background (CMB). The origin of the CMB itself is simple enough for most students to grasp, and the representation of the anisotropies using spherical harmonics is also manageable for undergraduate students who have previously encountered spherical harmonics in quantum mechanics courses. A more challenging aspect of developing familiarity with the CMB lies in understanding how the power spectra respond to changes in cosmological parameters. Intuition for this behavior can be developed either through a strong understanding of the physical processes before and around recombination (see, e.g., Ref.~\onlinecite{waynehututorial}) or by studying how the power spectra change when individual cosmological parameters are varied. A combination of these two approaches is often the most effective way to build intuition for CMB anisotropies. However, the latter approach has traditionally been limited by the availability of suitable parameter-varying animations or by the technical expertise required for students to run an Einstein--Boltzmann solver for many different parameter configurations.

CosmoSlider is designed to address this limitation by allowing students to quickly and interactively explore the effects of specific cosmological parameters on the angular power spectra. By providing an easy-to-use, hands-on tool, students can engage more actively in the process of building intuition without needing advanced technical skills. Simply memorizing the qualitative behavior of the spectra is, however, insufficient for a solid fundamental understanding of the anisotropies. CosmoSlider should therefore be used in conjunction with instruction on the underlying physical mechanisms that give rise to the observed changes.

\subsection{Plotting tool}
The layout of CosmoSlider’s plotting area is responsive to different screen sizes, and a limited degree of customization is provided. As a result, the tool can also be used to quickly generate clear and visually appealing plots for presentations or coursework, without requiring users to run an Einstein--Boltzmann solver and post-process the results themselves. Future updates will provide a higher level of customization along with an option for exporting the plot instead of relying on capturing screenshots. See section~\ref{sec:updates} for a detailed list on future updates to CosmoSlider.

\section{Scheduled updates}\label{sec:updates}
We plan to improve the functionality of both the iOS version and the web version in the near future and we are currently planning to include the following:

\begin{itemize}
    \item Introduce an option for saving the plotting area as an image. It should be possible for the user to specify the desired size and aspect ratio of the figure window such that the user is not limited by the device they are using.
    \item Provide more options for customization including colors of the graph and background as well as an option to overlay multiple graphs with different parameter configurations.
    \item Make more observables available to plot. This includes the matter power spectrum as well as background quantities.
    \item Allow points from more data sets to be plotted alongside the spectra. This could include additional CMB experiments such as SPT~\cite{Simard:2017xtw}, ACT~\cite{AtacamaCosmologyTelescope:2025blo}, WMAP~\cite{wmap}, etc. along with the option of importing a custom set of data points to use.
    \item Expanding the resources and informations available in the application by including more links to relevant reading material and similar. 
\end{itemize}

\noindent Both versions of CosmoSlider are actively maintained, and additional updates are implemented as needed or in response to feature requests from the user community.

\section{Conclusion}\label{sec:conclusion}
We have presented CosmoSlider, a lightweight interactive tool that enables real-time exploration of the cosmic microwave background for educational purposes. By using a TensorFlow Lite emulator rather than a fine grid of precomputed spectra, CosmoSlider responds rapidly to parameter changes, ensuring a smooth and versatile user experience even when multiple parameters are varied simultaneously. Both an iOS version and a web-based version are available, providing cross-platform support, and both are actively maintained to ensure long-term usability.

When used in conjunction with explanations of the underlying physical mechanisms responsible for CMB anisotropies, CosmoSlider can help students build intuition for how the CMB power spectra respond to changes in cosmological parameters. This makes the tool well suited for use in lectures, demonstrations, and self-guided exploration, where immediate visual feedback can reinforce theoretical concepts. CosmoSlider is not intended to replace detailed numerical calculations performed with Einstein--Boltzmann solvers, but rather to complement them by facilitating qualitative understanding and exploration.

In the near future, several aspects of CosmoSlider will be updated and refined, including customization options, additional data sets, and a broader set of emulated observables. More broadly, this work illustrates how modern emulation techniques can be used to make complex physical models more accessible in an educational context, suggesting similar approaches may be valuable in other areas of physics education.

\section*{Acknowledgments}
We acknowledge computing resources from the Centre for Scientific Computing Aarhus (CSCAA). A.N. was supported by the Carlsberg Foundation, grant CF24-1944.

\bibliographystyle{utcaps}
\bibliography{total}

\providecommand{\href}[2]{#2}\begingroup\raggedright\begin{thebibliography}{10}

\bibitem{Blas:2011rf}
D.~Blas, J.~Lesgourgues, and T.~Tram, ``{The Cosmic Linear Anisotropy Solving
  System (CLASS) II: Approximation schemes},''
  \href{http://dx.doi.org/10.1088/1475-7516/2011/07/034}{{\em JCAP} {\bfseries
  07} (2011)  034}, \href{http://arxiv.org/abs/1104.2933}{{\ttfamily
  arXiv:1104.2933 [astro-ph.CO]}}.

\bibitem{Lewis:1999bs}
A.~Lewis, A.~Challinor, and A.~Lasenby, ``{Efficient computation of CMB
  anisotropies in closed FRW models},''
  \href{http://dx.doi.org/10.1086/309179}{{\em Astrophys. J.} {\bfseries 538}
  (2000)  473--476}, \href{http://arxiv.org/abs/astro-ph/9911177}{{\ttfamily
  arXiv:astro-ph/9911177}}.

\bibitem{waynehututorial}
W.~Hu, ``{CMB Power Animations}.'' \\
  \url{https://background.uchicago.edu/~whu/metaanim.html}, 2004.
\newblock Accessed: 2026-01-13.

\bibitem{maxtegmarkuniverses}
M.~Tegmark, C.~Williams, and I.~Zelko, ``The cosmic flic.'' \\
  \url{https://space.mit.edu/home/tegmark/movies.html}, 2013.
\newblock Accessed: 2026-01-13.

\bibitem{Planck:2018vyg}
{\bfseries Planck} Collaboration, N.~Aghanim {\em et al.}, ``{Planck 2018
  results. VI. Cosmological parameters},''
  \href{http://dx.doi.org/10.1051/0004-6361/201833910}{{\em Astron. Astrophys.}
  {\bfseries 641} (2020)  A6},
  \href{http://arxiv.org/abs/1807.06209}{{\ttfamily arXiv:1807.06209
  [astro-ph.CO]}}. [Erratum: Astron.Astrophys. 652, C4 (2021)].

\bibitem{buildauniverse}
T.~W. Collaboration, ``{Build a Universe: CMB Analyzer}.'' \\
  \url{https://lambda.gsfc.nasa.gov/bau/}, 2013.
\newblock Accessed: 2026-01-13.

\bibitem{wmap}
G.~Hinshaw, D.~Larson, E.~Komatsu, D.~N. Spergel, C.~L. Bennett, J.~Dunkley,
  M.~R. Nolta, M.~Halpern, R.~S. Hill, N.~Odegard, L.~Page, K.~M. Smith, J.~L.
  Weiland, B.~Gold, N.~Jarosik, A.~Kogut, M.~Limon, S.~S. Meyer, G.~S. Tucker,
  E.~Wollack, and E.~L. Wright,
  \href{http://dx.doi.org/10.1088/0067-0049/208/2/19}{``NINE-YEAR WILKINSON
  MICROWAVE ANISOTROPY PROBE ( WMAP ) OBSERVATIONS: COSMOLOGICAL PARAMETER
  RESULTS,''{\em The Astrophysical Journal Supplement Series} {\bfseries 208}
  (Sept., 2013)  19}. \url{http://dx.doi.org/10.1088/0067-0049/208/2/19}.

\bibitem{nasalambda}
G.~E. Addison, T.~M. Essinger-Hileman, M.~R. Greason, T.~B. Griswold, T.~Jaffe,
  N.~Miller, U.~Prasad, and J.~L. Weiland, ``{{Legacy Archive for Microwave
  Background Data Analysis (LAMBDA): An Overview}},''
  \href{http://arxiv.org/abs/1905.08667}{{\ttfamily arXiv:1905.08667
  [astro-ph.IM]}}.

\bibitem{cosmowebapp}
J.~Helsby and E.~Baxter, ``{Interactive Cosmological Power Spectra}.'' \\
  \url{http://redshiftzero.com/cosmowebapp/}, 2017.
\newblock Accessed: 2026-01-13.

\bibitem{cmbsimulator}
S.~Lowe and C.~North, ``{Planck CMB Simulator}.'' \\
  \url{http://chrisnorth.github.io/planckapps/Simulator/}, 2013.
\newblock Accessed: 2026-01-13.

\bibitem{cmbverse}
G.~Montefalcone, C.~Enlowsmith, S.~Dunkerley, and J.~Chubick, ``{CMBverse}.''
  \url{https://gabrielemontefalcone.com/CMBverse/}, 2025.
\newblock Accessed: 2026-01-13.

\bibitem{webcamb}
LAMBDA, ``{CAMB Online}.'' \\
  \url{https://lambda.gsfc.nasa.gov/toolbox/camb_online.html}, 2005.
\newblock Accessed: 2026-01-13.

\bibitem{cmbexplorer}
S.~Chaulagain, ``{CMB Explorer}.''
  \url{https://cmb-explorer-camb.streamlit.app}, 2025.
\newblock Accessed: 2026-01-13.

\bibitem{rossjenkinson}
R.~Jenkinson, ``{Interactive CMB TT spectrum (CAMB)}.'' \\
  \url{https://rossjenkinson-cmb.streamlit.app}, 2025.
\newblock Accessed: 2026-01-13.

\bibitem{Nygaard:2022wri}
A.~Nygaard, E.~B. Holm, S.~Hannestad, and T.~Tram, ``{CONNECT: a neural network
  based framework for emulating cosmological observables and cosmological
  parameter inference},''
  \href{http://dx.doi.org/10.1088/1475-7516/2023/05/025}{{\em JCAP} {\bfseries
  05} (2023)  025}, \href{http://arxiv.org/abs/2205.15726}{{\ttfamily
  arXiv:2205.15726 [astro-ph.IM]}}.

\bibitem{Efstathiou:2003dj}
G.~Efstathiou, ``{Myths and truths concerning estimation of power spectra},''
  \href{http://dx.doi.org/10.1111/j.1365-2966.2004.07530.x}{{\em Mon. Not. Roy.
  Astron. Soc.} {\bfseries 349} (2004)  603},
  \href{http://arxiv.org/abs/astro-ph/0307515}{{\ttfamily
  arXiv:astro-ph/0307515}}.

\bibitem{Simard:2017xtw}
G.~Simard {\em et al.}, ``{Constraints on Cosmological Parameters from the
  Angular Power Spectrum of a Combined 2500 deg$^2$ SPT-SZ and Planck
  Gravitational Lensing Map},''
  \href{http://dx.doi.org/10.3847/1538-4357/aac264}{{\em Astrophys. J.}
  {\bfseries 860} (2018) no.~2, 137},
  \href{http://arxiv.org/abs/1712.07541}{{\ttfamily arXiv:1712.07541
  [astro-ph.CO]}}.

\bibitem{AtacamaCosmologyTelescope:2025blo}
{\bfseries Atacama Cosmology Telescope} Collaboration, T.~Louis {\em et al.},
  ``{The Atacama Cosmology Telescope: DR6 power spectra, likelihoods and
  {\ensuremath{\Lambda}}CDM parameters},''
  \href{http://dx.doi.org/10.1088/1475-7516/2025/11/062}{{\em JCAP} {\bfseries
  11} (2025)  062}, \href{http://arxiv.org/abs/2503.14452}{{\ttfamily
  arXiv:2503.14452 [astro-ph.CO]}}.

\end{thebibliography}\endgroup
\end{document}